\documentclass{INTERSPEECH2023}
\usepackage{blindtext}
\usepackage{hyperref}

\interspeechcameraready

\title{Exploring Domain-Specific Enhancements for a Neural Foley Synthesizer}
\name{ Ashwin Pillay$^{1*}$, Sage Betko$^{1*}$,  Ari Liloia$^{1}$, Hao Chen $^{1}$, Ankit Shah$^{1*}$ }
\address{
  $^1$Carnegie Mellon University, Pittsburgh, PA
}
\email{\{apillay, sbetko,  aliloia, haoc3, aps1\}@andrew.cmu.edu}

\begin{document}

\ninept
\maketitle

\def\thefootnote{*}\footnotetext{These authors contributed equally to this work}\def\thefootnote{\arabic{footnote}}

\begin{abstract}


Foley sound synthesis refers to the creation of authentic, diegetic sound effects for media, such as film or radio. In this study, we construct a neural Foley synthesizer capable of generating mono-audio clips across seven predefined categories. Our approach introduces multiple enhancements to existing models in the text-to-audio domain, with the goal of enriching the diversity and acoustic characteristics of the generated foleys. Notably, we utilize a pre-trained encoder that retains acoustical and musical attributes in intermediate embeddings, implement class-conditioning to enhance differentiability among foley classes in their intermediate representations, and devise an innovative transformer-based architecture for optimizing self-attention computations on very large inputs without compromising valuable information. Subsequent to implementation, we present intermediate outcomes that surpass the baseline, discuss practical challenges encountered in achieving optimal results, and outline potential pathways for further research. Note: This system was submitted to Task 7 of the DCASE 2023 challenge, and the relevant codebase can be accessed at: \url{https://github.com/ankitshah009/foley-sound-synthesis_DCASE_2023}.

\end{abstract}


\section{Introduction}

Foley sound refers to diegetic, non-musical sound effects that convey the sounds produced by events depicted in a piece of media, such as radio or film. The process of creating complex sound environments from scratch is time-consuming and expensive; a method for convincingly synthesizing sounds could improve the content creation workflow. It could also be used to synthesize and augment other datasets. In this project, we create a machine learning model that generates original audio clips belonging to one of seven foley sound categories, namely \textit{DogBark, Footstep, GunShot, Keyboard, MovingMotorVehicle, Rain}, and \textit{Sneeze/Cough}  \cite{Davis80-COP}. Evaluating present results, our system has exceeded the performance of the DCASE baseline model in six out of seven categories, as measured via Frechet Audio Distance (FAD).


\section{Literature Review}

Previous work by Ghose \& Provost \cite{ghose2020autofoley}, \textit{AutoFoley}, describes an ensemble of a CNN + Fast-Slow LSTM model and a CNN + Temporal Relation Network (TRN) to generate foleys for the provided silent video input. The model is trained on a novel dataset to generate several classes of foleys. This is done by predicting a sound class matrix and combining each component with the average spectrogram of the corresponding foley class to generate a final audio output for the given video frame.

Additionally, foley synthesis is a subset of the text-to-audio (TTA) generation problem that has received considerable deep-learning research attention in recent times. Kreuk et al. \cite{kreuk2022audiogen} developed \textit{Audiogen}, a TTA generator using a combination of autoregressive audio encoder-decoder and language transformer-decoder model that can outperform prior work in this field by Yang et al. \cite{yang2022diffsound}. Audiogen is trained end-to-end on a combination of input audio and a corresponding textual description. Internally, the audio and text are encoded into compressed representations for improving the speed and generalization of the model. While Audiogen can generate audio for text prompts it was not explicitly trained on, the resulting output may not follow the temporal token ordering of the input prompts.

Recently, \textit{AudioLDM} developed by Liu et al. \cite{liu2023audioldm} achieved improved results over Audiogen in terms of both subjective metrics and objective metrics such as Frechet Audio Distance (FAD). AudioLDM is a Latent Diffusion Model (LDM) based TTA generator that uses contrastive language-audio pretraining (CLAP) models \cite{elizalde2022clap} to represent audio-text cross-modalities and a Variational Autoencoder (VAE) + HiFi-GAN \cite{kong2020hifi} combination to synthesize audio from its latent space representation. Using CLAP enables the model to be trained on embeddings directly derived from audio, bypassing the intrinsic inefficiencies and human-induced inconsistencies of textual audio description. During inference, the text prompt provided is converted into its audio embedding by CLAP, and is subsequently converted into a latent audio representation by the LDM. While AudioLDM is a good reference for our research, it would be imperative to significantly optimize the model specific to fixed-class foley generation while being closer to the respective ground truth on subjective and objective evaluation metrics.

It is also appropriate to note the success of the three-stage DTFR model \cite{Liu_MLP2022_01}, which is the baseline model for the DCASE challenge and is explored in detail later in this report. 


\section{Model Description} \vspace{-2mm}

We began our work by reproducing the baseline provided by the DCASE2023 Task 7 organizers to recreate the stated results and identify strategies to improve upon them. Our work was originally concerned with making optimizations to the given baseline that enabled us to regenerate the provided results on a single GPU. Subsequently, we made enhancements to several components of the baseline with the goal of improving the quality and variety of the generated foleys.

\subsection{Optimizations to the baseline}
Upon experimenting with the baseline model, we observed that the learning rate for the VQ-VAE model was too large to yield any meaningful result, so we added a cyclic learning rate scheduler. Considering our time and compute constraints, we developed an optimized training scheme that could give us acceptable results within a days worth of training on a single, consumer-grade GPU. We accomplished this by implementing mixed precision training. We also ablated our batch size, reducing them to 16 for VQ-VAE and 8 for PixelSNAIL training. Lastly, we also implemented a system that employs the trained model on inference mode to return the FAD scores for 32 randomly-generated foleys of each of the aforementioned classes for subsequent evaluation.

\subsection{Using CEmbed: An enhanced audio representation}

The baseline model is trained on, and generates melspectrogram representations of foley audios. Specifically, it uses 80 mel filter banks, an FFT size of 1024 and a hop size of 256 to obtain the melspectrogram. This converts 4s of foleys sampled at 22050 Hz to 80x344 vectors, ie, a  $\approx 3.2$x compression of data. We speculate that this compression undergoes significant compromises in acoustical and spectral information that could have improved the quality and accuracy of the underlying statistical distributions our downstream model approximates each foley class to.

As an alternative, we propose enhancing the melspectrogram input with higher-level audio features corresponding to factors like its key and acoustics. We believe such representations aid the model to utilize more domain-specific information while learning intra-class and inter-class qualities of the foleys. To this end, we integrated a pretrained encoder,  MERT-v1-330M, as a preprocessor to our system.

MERT \cite{li2022large} \footnote{MERT-v1-330M Huggingface: \url{https://huggingface.co/m-a-p/MERT-v1-330M}} is a large-scale model trained on music audio for general music understanding. It has an architecture similar to HuBERT\cite{hsu2021hubert}, a model for self-supervised speech representation learning  that has been proven to capture higher-level acoustical features than melspectrograms. While HuBERT is trained on 16 kHz speech data, MERT has been specifically trained using a Masked Language Model (MLM) paradigm on 24 kHz music / audio data. The audio-specificity of MERT embeddings and its higher sampling rate results in more granular and meaningful representation of foley features than HuBERT embeddings. Moreover, MERT has been validated against a variety of music information retrieval (MIR) tasks like genre classification and key detection. The developers of MERT state that across the zeroth dimension of its embeddings, there is a gradual increase in the level of features, e.g. the first few dim0 features represent lower-level features like the time-frequency variations and the last few represent higher-level features like the key to which the piece of input audio belongs. While features like the key are more relevant to music than foleys, we believe the model could utilize this information to identify differences between foleys of the same class; for e.g. differences in the bark of a young Chihuahua and an adult Bulldog.

To aid concatenation of the melspectrograms with MERT embeddings, we modified how the former was obtained. This was done by increasing the mel frequency bands to 129 and increasing the hop size to 320 samples. We hypothesize that the increased features provided by MERT will compensate for the increase in melspectrogram hop size. Finally, we combined the two embeddings to form  Combined Embeddings (\textbf{``CEmbed"}), as shown in Fig 1.

The use of CEmbed over plain melspectrograms required retraining all the downstream models in our system, along with significant changes to their architectures as described in the following subsections. For a brief comparison of the changes made to the input embedding of the baseline and the final model, refer Table 1.

\begin{table}[h!]
  \begin{center}
    \caption{Differences in sizes between analagous variables used in the baseline and final models. The melspectrogram sizes are (frequency band step, time step).}
    \label{tab:table1}
    \begin{tabular}{l|c|r} 
      Variable & Baseline Shape & Final Shape\\
      \hline
      Audio Input & (22050 x 4, 1) & (24000 x 4, 1)\\
      Melspectrogram & (80, 344) & (129, 300)\\
      MERT Encodings & - & (1023, 300)\\
      Input to VQVAE & (80, 344) & (1152, 300)\\
      VQ-VAE Latent & (20, 80) & (288, 75)\\
    \end{tabular}
  \end{center}
\end{table}

\begin{figure}
  \centering
  \includegraphics[width=\linewidth]{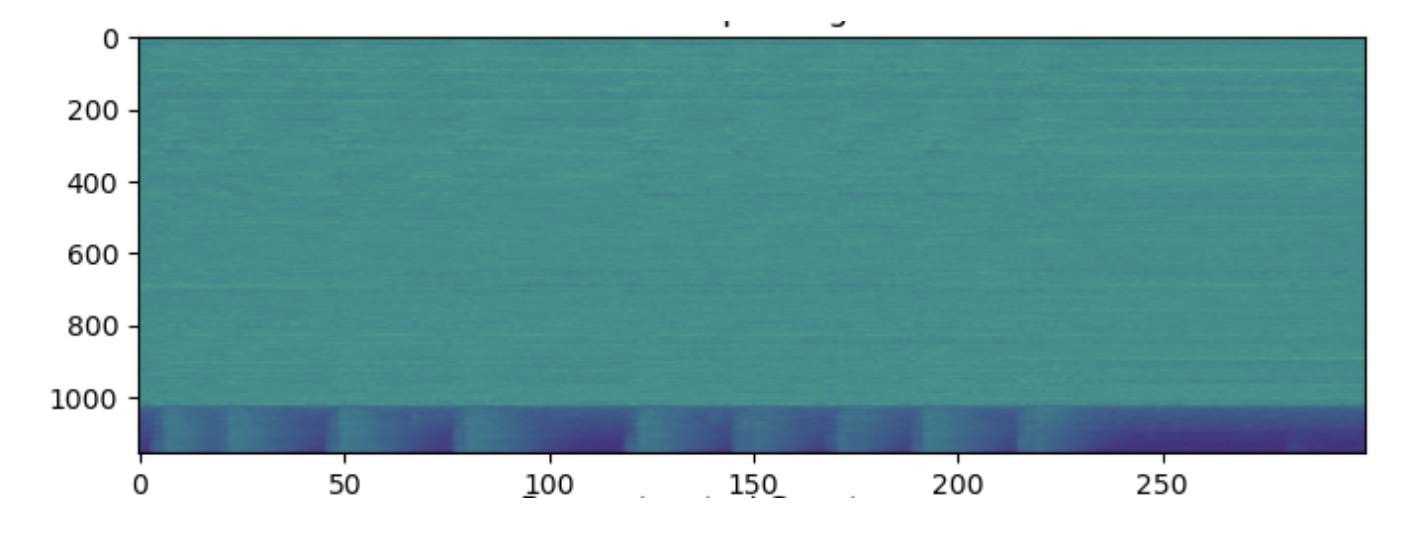}
  \caption{Plot of a CEmbedding for one sample in the development set. The lowest and least uniform-looking rows represent the melspectrogram, while the upper rows are made up of the MERT-generated embeddings.}
  \label{fig:multi_cembed}
\end{figure}

\subsection{VQ-VAE} \vspace{-2mm}
A latent variable model works under the assumption that given a  vector of latent variables $z$ and a dataset with data points $x$, the model can closely approximate $x$ using different values of $z$. Formally, we wish to optimize some vector $\theta$ in some space defined by the dimensions of $z$ and $x$ such that the probability of generating each $x$ in the dataset is maximized, according to 
\begin{equation}
    p(x) = \int p(x|z;\theta)p(z)dz
\end{equation}
where $p(x|z;\theta)$ is a Gaussian distribution, such that optimization techniques can be used to increase $p(x)$. 
A variational autoencoder (VAE) attempts to calculate $p(x)$ only based on the values of $z$ which are most likely to have produced $x$. We define a posterior categorical distribution $q(z|x)$ that gives the distribution over the values of $z$ likely to produce $x$. \cite{Doersch16-AV} These function make up a VAE, which consists of an encoder network that parameterizes $q(z|x)$, a prior distribution $p(z)$, and a decoder with a distribution $p(x|z)$ over input data.

\begin{align}
    z_{q}(x) = e_{k} \\
    k = \arg\min_{j}||z_{e}(x) - e_{j}||_{2}
\end{align}
This estimator is used to calculate the reconstruction loss, $\log(p(x|z_{q}(x))$. The gradient for this function can be approximated by the gradients from the decoder input; however, making this approximation effectively bypasses the embeddings during backpropagation, so a different method is necessary to learn the codebook. \cite{van_den_oord_neural_2017} For this task, the Vector Quantization (VQ) algorithm is used to form a quantized approximation to a distribution of input data vectors using a finite number of codebook vectors, then uses the Euclidean distance between them to adjust the latter toward the former. This results in a VQ loss term, $||\textnormal{sg}[z_{e}(x)]-e||^{2}_{2}$, where sg denotes the stopgradient operator, which detaches its argument from the computational graph. The volume of the embedding space is not constrained, so it is necessary to add another loss term to prioritize committing to an existing embedding over adding a new $e_{i}$ to the codebook. This term is $\beta||z_{e}(x) - \textnormal{sg}[e]||_{2}^{2}$, where $\beta$ is a tunable hyperparameter.
The full loss function for the VQ-VAE is then 
\begin{equation}
    L = \log(p(x|z_{q}(x)) + ||\textnormal{sg}[z_{e}(x)]-e||^{2}_{2} + \beta||z_{e}(x) - \textnormal{sg}[e]||_{2}^{2}
\end{equation}
\cite{van_den_oord_neural_2017} Within the baseline implementation provided by the DCASE organizers, a VQ-VAE model is used to learn a discrete-time frequency representation of the sounds in the training dataset.

\subsection{Enhancements to VQ-VAE: MVQVAE}

To ensure that the latent representations of foleys generated from the incoming CEmbeds utilize as much of the useful information as efficiently as possible, we proposed several changes to the baseline VQVAE architecture. The resulting model is termed MERT - VQVAE (\textbf{MVQVAE}) with its main enhancements described as follows:

\subsubsection{Foley Conditioning}\label{sec:mvqvae_conditioning}

The baseline VQ-VAE model learns an unconditional representation of sound, without any additional information about the category of sound during optimization or inference. Hence the responsibility of conditional sound generation lies solely with PixelSNAIL, which is tasked with learning to sample from the generalized codewords that make of the VQ-VAE's codebook, in order to assemble sequences based on the unique distribution of each sound category. However, the baseline VQ-VAE tends to produce codewords with similar conditional distributions across foley categories, which can make it difficult for PixelSNAIL to learn category-specific distributions. We hypothesize that this difficulty arises because the similar distributions cause PixelSNAIL to confuse categories, resulting in poor generation quality. To address this, we introduce a single linear layer that receives the average pre-quantization channel values of the latent representation and predicts the foley category to which the input belongs. The cross-entropy loss between the predicted and the actual foley category is added to the total loss scaled by a factor of $1 \times 10^{-2}$.

\subsubsection{CEmbed-specific model expansions}

Since the CEmbeddings in our new model are $\approx 14$ times larger than the melspectrograms, the baseline VQVAE cannot operate on them as is. Thus, one key enhancement brought by MVQVAE include increasing the size of the dictionary maintaining the codebook vectors that can represent a single encoder output from 512 to 1024. Additionally, we added a parallel ResNet block in the encoder and decoder to increase its capacity to grasp the increased information provided by the CEmbeds. Further, we included asynchronous time and frequency-masking data augmentations in the training paradigm to prevent the model from over-associating redundant relationships that may exist in the melspectrogram and the MERT embeddding of a given CEmbed. Fig 2 demonstrates this training paradigm.

\begin{figure}
  \centering
  \includegraphics[width=\linewidth]{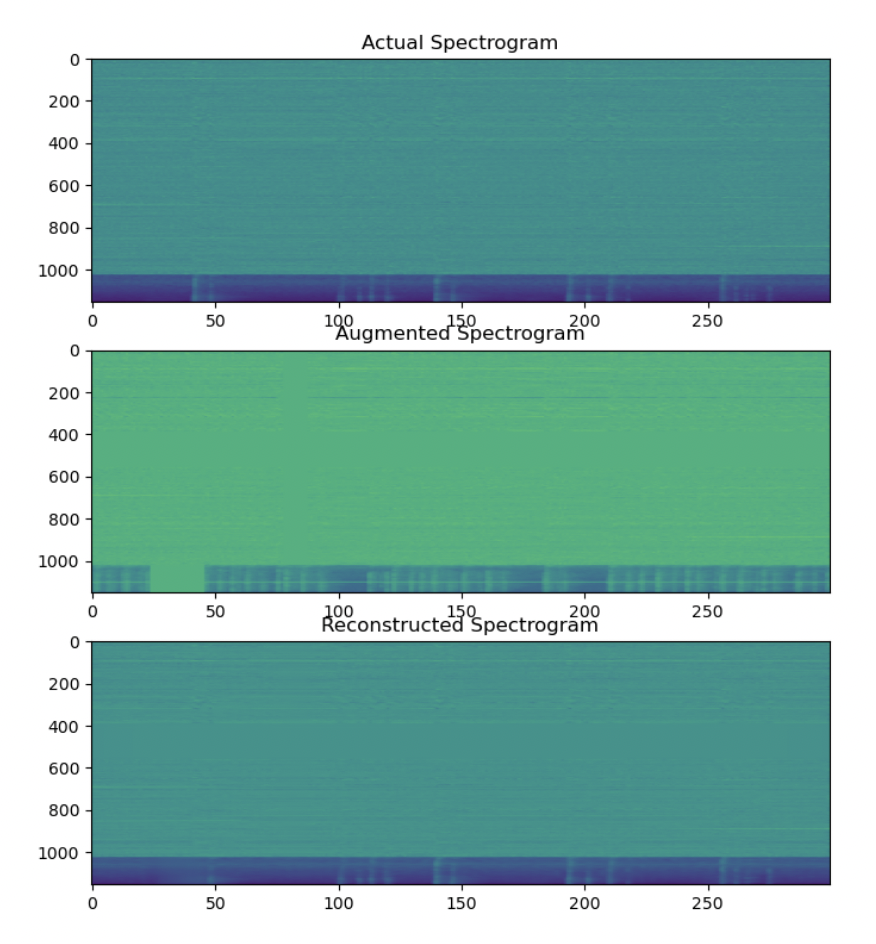}
  \caption{From top to bottom: the actual CEmbedding, an example of an augmented training input to the model, and the reconstructed output of MVQVAE}
  \label{fig:cembed2}
\end{figure}

\subsection{PixelSNAIL} \vspace{-2mm}
As mentioned in the previous section, generative models estimate the p(x), or the probability of observing some trait x. Autoregressive models factor the joint distribution as a product of conditionals over each feature.
\begin{equation}
p(x) = p(x_{1}, \ldots x_{n}) = \prod_{i=1}^{n}p(x_{i}|x_{1},\ldots x_{i-1})
\end{equation}
Autoregressive models implemented using traditional RNNs generally under-perform, possibly due to the temporally linear dependency of the information kept within hidden states from one time step to the next. Other architectures that would allow a model to easily refer to earlier parts of an input sequence are causal convolutions, which allow high bandwidth access over a finite context size, and self-attention models, which convert an input sequence into a set of key-value pairs, allowing access to an infinitely large context with low bandwidth access. The SNAIL method combines the two approaches by using the convolutions to aggregate information over which to build context and perform an attentive lookup. \cite{mishra2018simple} PixelSNAIL applies this strategy to autoregressive models. PixelSNAIL is comprised of residual blocks that carry out causal convolutions and attention blocks that produce keys and values from input data. \cite{chen2017pixelsnail} Within the baseline model provided by the DCASE organizers, PixelSNAIL is trained to learn the joint distribution of the discrete time frequency representation (DTFR) conditional on class label in order to autoregressively generate DTFR components. 

\subsubsection{Optimizing PixelSNAIL for CEmbed: Zen Mode} \vspace{-2mm}

When applied to CEmbeddings, the baseline version of PixelSNAIL suffers from impractical matrix multiplications. The scaled dot-product attention used in PixelSNAIL has an $O((TF)^2)$ memory requirement, where $T$ and $F$ are the time and feature dimensions of the quantized encodings from MVQVAE. This quadratic scaling makes self-attention impractical for longer sequence lengths, especially with the increased feature dimensionality introduced by MERT.  Our group proposed an approach called \textbf{Zen Mode} to balance PixelSNAIL's efficiency with the preservation of CEmbeddings' additional dimensionality.

Zen mode reduces the computation complexity of the self-attention mechanism in PixelSNAIL by incorporating trainable strided causal convolutional layers over the key and query vectors and transposed causal convolutions over the attention output. The convolutional layers downsample the input to the attention block, representing higher-level, coarser information from the embeddings and decreasing computational complexity. Our model applies a downsampling factor of 4, reducing the cost of computing the self-attention matrix by a factor of 16. Meanwhile, the actual CEmbed data is used without any downsampling. This allows us to model longer sequences while not sacrificing useful CEmbedding feature data in PixelSNAIL's decoder hidden states.

In the context of autoregressive models like PixelSNAIL, maintaining causality is essential. Standard transposed convolutions do not inherently possess causal properties, To address this, we introduce a novel technique called \textit{causal transposed convolution}. Causal transposed convolutions combine the upsampling capability of transposed convolutions with the causality property required for autoregressive modeling. This ensures that the generated output maintains causality, preserving the autoregressive nature of PixelSNAIL.

To the best of our knowledge, the use of zen mode and causal transposed convolutions have not yet been proposed in the machine learning literature, making this a unique contribution of this work. With these enhancements, we term the new model as \textbf{Zen PixelSNAIL}.

\newcommand*{\figuretitle}[1]{%
    {\centering
    \textbf{#1}
    \par\medskip}
}

\begin{figure}
    \centering
    \figuretitle{Melspectrograms during HiFi-GAN training, time vs. frequency (epochs 1, 94, 188)}
    \includegraphics[scale=0.3]{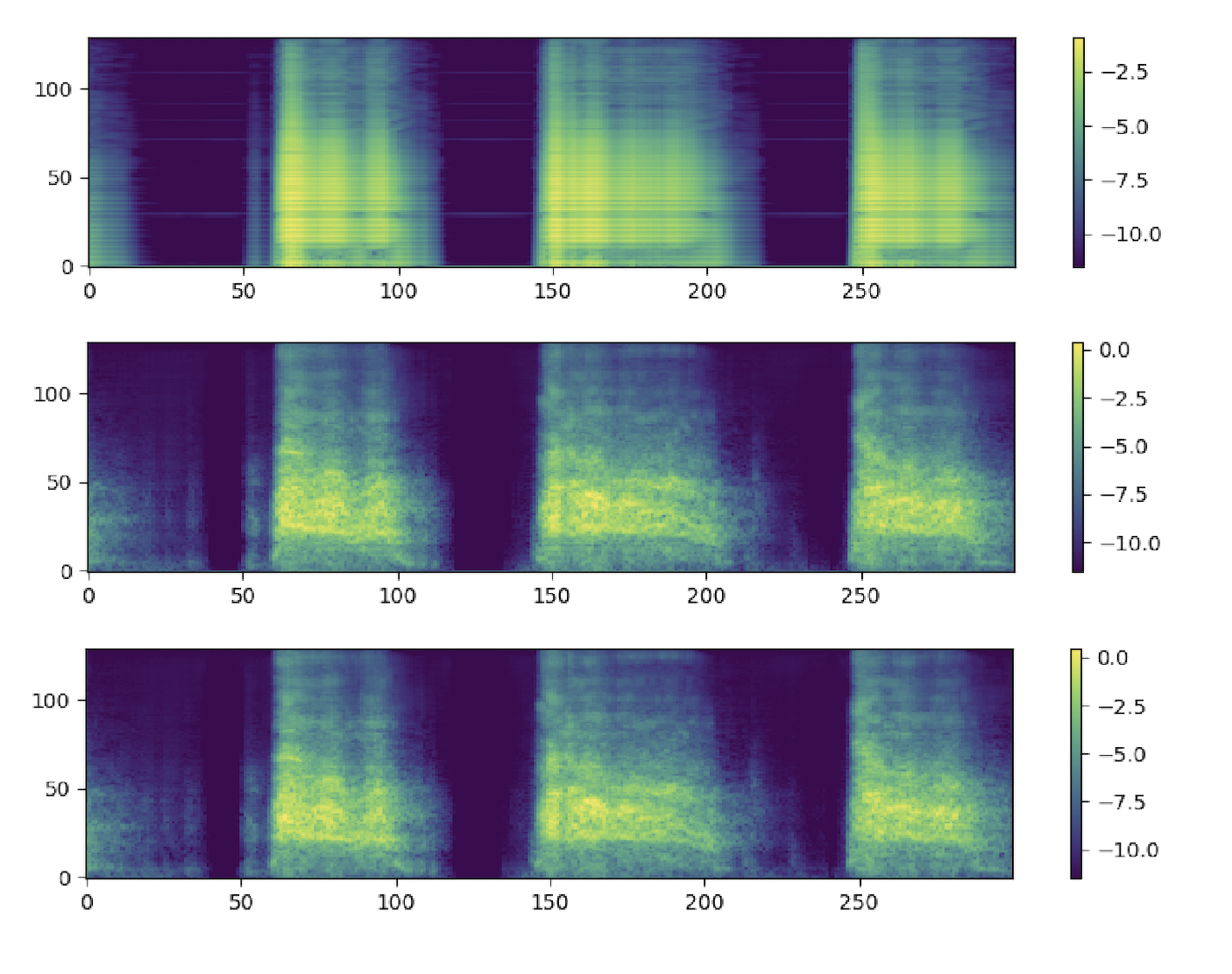}
    \caption{Frequency against time melspectrogram output of HiFi-GAN during training, at epochs 1, 94, and 188 (top to bottom) - over multiple epochs, the melspectrograms become more refined}
\end{figure}

\subsection{Modifications to HiFi-GAN: MHiFiGAN}

The pre-trained HiFi-GAN provided by the challenge organizers expects VQVAE-decoded melspectrograms to generate audio at 22050 Hz. Since MVQVAE returns decoded CEmbeds, we propose MERT HiFiGAN (\textbf{MHiFiGAN}), a model trained from scratch to vocode CEmbeds to audio at 24000 Hz. In contrast from HiFiGAN that performed dilation by a factor of 256, MHiFiGAN dilates incoming CEmbeds, which have a feature rate of 75 Hz, by a factor of 320. This also accounts for errors in rounding the duration of the foley sounds to 4 seconds, an area in which the previous model was prone to error.

To make MHiFiGAN robust against imperfections in the MVQVAE-decoded CEmbeds, we modified the training paradigm of MHiFiGAN such that its trained on time and frequency masked CEmbeds.

\section{Evaluation Metrics}

Our model will be evaluated on the quality of its output, which will be evaluated quantitatively via Frechet Audio Distance (FAD) and qualitatively via a subjective test.

FAD is an adaptation of the Frechet Inception Distance (FID) from the visual to the audio domain. Embedding statistics are extracted from a full evaluation set and a training set using VGGish, a pre-trained audio classification model. Multivariate Gaussians $\mathcal{N}_{e}(\mu_{e},\Sigma_{e})$ and $\mathcal{N}_{b}(\mu_{b},\Sigma_{b})$ are then computed on the evaluation and training sets. The Frechet distance between two Gaussians, 
\begin{equation}
	\textbf{F}(\mathcal{N}_{b}, \mathcal{N}_{e}) = ||\mu_{b} - \mu_{e}||^{2} + tr(\Sigma_{b} + \Sigma_{e} - 2\sqrt{\Sigma_{b}\Sigma_{e}})    
\end{equation}
is called the FAD score \cite{Kilgour19-AS}. FAD does not require a piece of reference audio to evaluate input, making it well-suited to evaluate this problem, as there is no specific ground truth for a clip falling into one of the categories. 

The subjective test will be carried out by the challenge organizers and members of other submission teams. Evaluators will judge the similarity between audio clips generated using the baseline model, audio clips generated using submitted models, and non-synthesized audio clips. Both fidelity and the degree to which the generated sound suits a category will be considered \cite{Davis80-COP}.

\section{Development Set}  \vspace{-2mm}

The development dataset provided by the DCASE organizers consists of 4,850 mono 16-bit 22,050 Hz sound clips from the UrbanSound8K, FSD50K, and BBC Sound Effects datasets. Each sound clip is exactly four seconds long and belongs to one of seven categories: DogBark, Footstep, GunShot, Keyboard, MovingMotorVehicle, Rain, and Sneeze/Cough. Per the challenge regulations, additional samples from these datasets are not permitted for training the foley synthesis system.

\begin{table}[h]
	\centering
	\caption{The number of foleys belonging to each category in the development set and their class ID.}
	\label{tab:num_cat}
	\begin{tabular}{|c|c|c|c|}
		\hline
		\textbf{ID} & \textbf{Category}  & \textbf{Number of Files} \\ \hline
		0                 & DogBark            & 617       \\ \hline
		1                 & Footstep           & 703       \\ \hline
		2                 & GunShot            & 777        \\ \hline
		3                 & Keyboard           & 800       \\ \hline
		4                 & MovingMotorVehicle & 581       \\ \hline
		5                 & Rain               & 741       \\ \hline
		6                 & Sneeze/Cough       & 631       \\ \hline
	\end{tabular}
\end{table}

\section{Preliminary Results}  \vspace{-2mm}

The DCASE development dataset was split into a train and validation set for model evaluation. The train set consisted of 4360 samples and the validation set contained 245 samples. The validation set was constructed with a stratified random sample where 35 samples were randomly selected from each category, and the remaining samples were assigned for training.

\subsection{Baseline Model}

We have successfully implemented and trained the baseline solution described in \cite{Liu_MLP2022_01}, surpassing the results of the challenge organizers in all seven foley sound categories. The baseline model's FAD scores evaluated on the DCASE development dataset are provided in Table \ref{tab:baseline-fad}.


\begin{table}[h]
	\centering
	\caption{FAD scores on DCASE development set (lower is better).}
	\label{tab:baseline-fad}
	\begin{tabular}{|c|c|c|c|}
		\hline
		\textbf{ID} & \textbf{Category}  & \textbf{FAD (DCASE)} & \textbf{FAD (Ours)} \\ \hline
		0                 & DogBark            & 13.411       & \textbf{8.958} \\ \hline
		1                 & Footstep           & 8.109        & \textbf{4.189 } \\ \hline
		2                 & GunShot            & 7.951        & \textbf{6.765 } \\ \hline
		3                 & Keyboard           & 5.230        & \textbf{3.086 }\\ \hline
		4                 & MovingMotorVehicle & 16.108       & \textbf{11.319} \\ \hline
		5                 & Rain               & 13.337       & \textbf{9.321 }\\ \hline
		6                 & Sneeze/Cough       & 3.770        & \textbf{2.675 }\\ \hline
	\end{tabular}
\end{table}

All code is available on our project GitHub repository, as noted in the abstract. This includes our implementation of the baseline model and the FAD computation. Our training runs for VQ-VAE and PixelSNAIL are openly available to view on Weights \& Biases.\footnote{\url{https://wandb.ai/audio-idl/Foley-sound-synthesis_DCASE_2023-baseline_dcase2023_task7_baseline}} We would also like to present a few example sounds generated by our current model in each category.\footnote{Audio synthesis examples: \url{https://drive.google.com/drive/folders/10LdqxEeVerVNEqcAb3uWjjpxnlmH27Jd}}

Following the training procedure by \cite{Liu_MLP2022_01}, we trained the VQ-VAE for 800 epochs with a learning rate of $3\times 10^{-3}$, although we reduced the batch size to 16 from 64 in order to fit within a single GPU. Notably, however, we have exceeded the baseline performance with only 265 training epochs of PixelSNAIL, whereas \cite{Liu_MLP2022_01} train for 1500 epochs. We attribute this improvement in efficiency primarily to our reduction in batch size from 32 to 8 and our addition of a cyclic learning rate scheduler with a reduced initial learning rate of $1\times 10^{-5}$. Our use of PyTorch's automatic mixed-precision (AMP) training enabled us to complete the training of both the baseline VQ-VAE and PixelSNAIL models in under 24 hours on a single NVIDIA RTX A4000 with 16GB of VRAM.

\subsection{Conditioned VQ-VAE and MVQVAE}

Table~\ref{tab:cvqvae_ablation_table} presents the results for the baseline and conditioned VQ-VAE models trained on Melspectrograms. Table~\ref{tab:cmvqvae_ablation_table} shows the same but for the MVQVAE. The addition of the classification loss term reduces both the train and validation MSE reconstruction loss.

Most notably, we see an extremely significant reduction in latent loss, which measures the difference between the pre- and post-quantization encodings. Since  the encoder output is mapped once to the codewords to obtain training data for PixelSNAIL, and then again to decode PixelSNAIL generation output during synthesis, it is critical to obtain a low latent loss. This measures the degree of misalignment between the codebook and the encoder output, and hence the level of noise introduced by mapping between the encoding and the latent codes.\cite{Liu_MLP2022_01}

We hypothesize that the addition of class-conditioning described in section \ref{sec:mvqvae_conditioning} while training the VQ-VAE/MVQVAE helps to better structure the latent space, as it allows the model to separate features unique to each sound category. This separation enables the codebook to hold more meaningful codewords that cater to individual sound categories, ultimately leading to a more effective use of the codebook's capacity.

\begin{table}[h!]
  \begin{center}
    \caption{Loss terms in the baseline (unconditioned) and conditioned Melspectrogram based VQ-VAE.}
    \label{tab:cvqvae_ablation_table}
    \begin{tabular}{|c|c|c|c|}
        \multicolumn{4}{c}{Train} \\
        \hline
        Model & MSE & Cross-Entropy & Latent Diff \\
        \hline
        Conditioned & \textbf{0.14056} & 0.02053 & \textbf{0.00167} \\
        Baseline & 0.14395 & -- & 0.00183 \\
        \hline
        \multicolumn{4}{c}{Validation} \\
        \hline
        Model & MSE & Cross-Entropy & Latent Diff \\
        \hline
        Conditioned & \textbf{0.14814} & 0.02145 & \textbf{0.00171} \\
        Baseline & 0.19166 & -- & 0.00222 \\
        \hline
    \end{tabular}
  \end{center}
\end{table}
\begin{table}[h!]
  \begin{center}
    \caption{Loss terms in the conditioned and unconditioned MVQVAE.}
    \label{tab:cmvqvae_ablation_table}
    \begin{tabular}{|c|c|c|c|}
        \multicolumn{4}{c}{Train} \\
        \hline
        Model & MSE & Cross-Entropy & Latent Diff \\
        \hline
        Conditioned &  \textbf{0.357} & 0.0859 & \textbf{0.0179} \\
        Unconditioned & 0.4084 & -- & 0.2973 \\
        \hline
        \multicolumn{4}{c}{Validation} \\
        \hline
        Model & MSE & Cross-Entropy & Latent Diff \\
        \hline
        Conditioned & \textbf{0.2636} & 0.02145 & \textbf{0.0208} \\
        Unconditioned & 0.3196 & -- & 0.3669 \\
        \hline
    \end{tabular}
  \end{center}
\end{table}

\subsection{MHiFi-GAN}
Since the baseline HiFiGAN model was pretrained and provided to us, we are unable to report its metrics to compare it with the results of MHiFi-GAN. However, through playback of the audio generated, we can validate that the model improves the quality of CEmbed to audio conversion over several epochs. Table 6 summarizes the validation and training metrics obtained for MHiFi-GAN after training it for 180 epochs. 


\begin{table}[h!]
  \begin{center}
    \caption{Training \& Validation Metrics for MHiFiGAN.}
    \label{tab:cmvqvae_ablation_table}
    \begin{tabular}{|c|c|c|}
        \hline
        \multicolumn{3}{|c|}{Train} \\
        \hline
        Discriminator Loss & Generator  Loss & Mel Recon. L1\\
        \hline
        3.041 &  27.911 & 0.3336 \\
        \hline
        \multicolumn{3}{|c|}{Validation} \\
        \hline
        Discriminator Loss & Generator Loss & Mel Recon. L1 \\
        \hline
        2.961 & 27.760 & 0.3281 \\
        \hline
    \end{tabular}
  \end{center}
\end{table}

\section{Obstacles to Final Results}


In our research, we propose a solution that consists of a cascade of three large models. Due to upstream modifications made to accommodate more detailed input representations, we had to enhance and train these models ourselves. One of the main challenges we faced during the training process was the requirement of having a fully trained MVQVAE to extract codes for Zen PixelSNAIL training. Despite implementing Zen mode optimizations, the increased size of Zen PixelSNAIL introduced numerous engineering challenges that impeded training.

We experimented with a few different MVQVAE configurations. The first of these, which we called MVQVAEv1, contained 512 codewords - the same number as the baseline Melspectrogram-based VQ-VAE. To train Zen PixelSNAIL on MVQVAEv1 codes and include CEmbeddings within our fixed compute budget of 16GB VRAM, we decreased its parameter count by reducing the number of channels from 256 to 128 and the number of residual blocks from 4 to 3. However, after several days, the model reached a saturation point at 50\% accuracy and could not learn further, necessitating training from scratch on a larger model.

Concurrently, we discovered that the 512-codeword limitation of MVQVAEv1 hindered its ability to reconstruct CEmbeddings. Consequently, we trained a second model, MVQVAEv2, with 1024 codewords, which resulted in better reconstruction MSE and significantly improved qualitative reconstruction during listening tests on the HiFi-GAN waveform output. Subsequently, we restored the channel count and the number of residual blocks in Zen PixelSNAIL to their original values and trained on the larger MVQVAEv2 codes.

Our final configuration, which is currently being trained on four NVIDIA A40 (48GB) GPUs, faced a multitude of engineering challenges as we attempted to scale up. The larger model was particularly susceptible to exploding gradients, which corrupted the optimizer state. Due to Zen PixelSNAIL's four serial decoder blocks, a significant accumulation of error occurred when applied to the larger CEmbeddings. To stabilize training, we implemented gradient clipping and experimented with different values of the maximum gradient norm. The training is currently ongoing, and we hope to achieve further advancements with this configuration.

Once Zen PixelSNAIL is sufficiently trained, we expect the overall system to be able to generate the specified number of foleys of each class with the fidelity and variety of each foley being considerable better than the baseline. We intend to validate the same using the evaluation strategies described in Section 4.

\section{Conclusion}

In our work, we aim to develop a neural sound synthesis engine capable of generating foleys belonging to predefined classes. Our goal is for the generated sounds to exhibit higher quality (comparable to human-generated foleys in a studio) and increased variety compared to general-purpose text-to-audio models and existing baselines. To achieve this, we create embeddings that represent both the lower-level time-frequency variances and the higher-level acoustical and musical features of the foleys. We then enhance our models to utilize this information for the intrinsic development of more detailed and distinguishable statistical distributions of each foley class.

Regarding model improvements, we introduced potentially innovative techniques such as class conditioning to increase the inter-class distance between foleys, Zen Mode to streamline attention-context computations without sacrificing input quality, and Causal Transpose CNNs to support dilation in auto-regressive prediction problems.


\subsection{Future Work}
\begin{enumerate}
	\item \textbf{Reducing the dimensions of MVQVAE latent encodings}: To alleviate Zen PixelSNAIL training, a simple strategy would be to modify MVQVAE ResNet's to output encodings of lower dimensionality at a lower vector rate. Another modification would be to add a CNN layer to compress MERT embeddings effectively. However, a major challenge in this case would be to identify the right tradeoff between granularity of information and computational load.
    \item \textbf{Identifying alternatives to Zen PixelSNAIL}: It would be logical to identify complete alternatives to autoregressive approaches like PixelSNAIL. Diffusion models used in works like AudioLDM would be a popular option to consider in this case.
    \item \textbf{Identifying alternatives to MVQVAE}: Alternatives like improved VQ-Diffusion models \cite{tang2023improved} may eliminate the unidirectional bias and accumulation of errors of the auto regressive approach, thus avoiding the quadratic attention cost of PixelSNAIL.
\end{enumerate}

\bibliographystyle{IEEEtran}
\bibliography{mybib}

\end{document}